\documentstyle[12pt]{article}

\begin{document}
\baselineskip 24pt
\begin{center}
\baselineskip 24pt
\noindent
{\large\bf THE TOPOLOGICAL STRUCTURE OF THE VORTICES\\
IN THE $O(n)$ SYMMETRIC TDGL MODEL$^*$\footnote{$^*$ Work supported by the
National Nature Science Foundation of China.}}

\vskip 24pt
\normalsize
\noindent
YISHI DUAN, YING JIANG$^\dagger$\footnote{$^\dagger$ Corresponding author;
E-mail: itp3@lzu.edu.cn} and TAO XU\\
{\it Institute of Theoretical Physics, Lanzhou University, 
Lanzhou, 730000, P.R.China}

\vskip 48pt
\begin{minipage}{137mm}
\baselineskip 24pt
\normalsize
\noindent 
In the light of $\phi $--mapping method and topological current theory, the
topological structure of the vortex state in TDGL model and the topological
quantization of the vortex topological charges are investigated. It is pointed
out that the topological charges of the vortices in TDGL model are described
by the Winding numbers of $\phi $--mapping which are determined in terms of
the Hopf indices and the Brouwer degrees of $\phi $--mapping.
\end{minipage}
\end{center}

\vskip 48pt
\noindent 
The world of defects is amazingly rich and have been the focus of much
attention in many areas of contemporary physics$^{\cite{cos}-\cite{gue}}$. The
importance of the role of defects in understanding a variety of problems in
physics is clear$^{\cite{jan}-\cite{cen}}$. As is well known, topology now
becomes much more important and necessary in physics, hence it is necessary
for us to investigate the topological properties of the defects
meticulously. In our previous work, by the use of the $\phi $--mapping
topological current theory$^{\cite{phimap}}$, we have investigated the
topological invariants$^{\cite{lee}-\cite{li}}$ and the topological structures of
physical systems$^{\cite{14}-\cite{zhh}}$ successfully. Now, in the light of this
useful method, we will study the topological properties of the vortices in
the context of an $O(n)$ symmetric time--dependent Ginzburg--Landau (TDGL)
model for the case of point defects$^{\cite{pointdef}}$ where $n=k$ and $k$ is
the spatial dimensionality.

We consider a time-dependent Ginzburg--Landau model for an $n$--component
order parameter $\vec \phi (\vec r,t)=(\phi ^1(\vec r,t),\cdots ,\phi ^n(%
\vec r,t))$ governed by the Langevin equation 
\begin{equation}
\label{00}\frac{\partial \vec \phi }{\partial t}=\vec K=-\Gamma \frac{\delta
F}{\delta \vec \phi }+\vec \eta 
\end{equation}
where $\Gamma $ is a kinetic coefficient and $\vec \eta $ is a thermal noise
which is related to $\Gamma $ by a fluctuation--dissipation theorem. $F$ is
a Ginzburg--Landau effective free energy assumed to be of the form 
\begin{equation}
\label{000}F=\int d^kr\left[ \frac c2(\nabla \vec \phi )^2+V(\vec \phi
)\right] 
\end{equation}
where $c<0$ and the potential $V$ is assumed to be of the degenerate
double--well form.

In recent years, many works have been done on the system of TDGL model. Liu
and Mazenko have discussed the growth kinetics of the systems with
continuous symmetry$^{\cite{maz1}}$, the defect--defect correlation in the
dynamics of first--order phase transitions$^{\cite{maz2}}$ and the vortex
velocities in the TDGL model$^{\cite{maz3}}$. Ryusuke Ikeda has presented the
hydrodynamical description for vortex states in type II superconductors on
the TDGL equation$^{\cite{ryu}}$. By the use of the TDGL model, Sch\"onborn and
Desai have investigated the intra--surface kinetics of phase ordering on
toroidal and corrugated surfaces$^{\cite{schon}}$. Two--dimensional $XY$ models
with resistively--shunted junction dynamics have also been discussed by Kim
et.al$^{\cite{kim}}$. However, most of them concentrated on the dynamical
properties of the TDGL model. In this letter, we will focus on the intrinsic
topological structure of the vortex topological current and give the
topological quantization of the topological charges of the vortices in TDGL
model.

It is well known that the $n$--dimensional order parameter field $\vec \phi (%
\vec r,t)$, which is governed by the Langevin equation, determines the
defect properties of the system, and it can be looked upon as a smooth
mapping between the $(n+1)$--dimensional space--time $X$ and the $n$%
--dimensional Euclidean space $R^n$ as $\phi :X\rightarrow R^n$. By analogy
with the discussion in our previous work$^{\cite{zhh},\cite{11}}$, from this $\phi $--mapping, a
topological current can be deduced as 
\begin{equation}
\label{1}j^\mu =\frac 1{A(S^{n-1})(n-1)!}\epsilon ^{\mu \mu _1\cdots \mu
_n}\epsilon _{a_1\cdots a_n}\partial _{\mu _1}n^{a_1}\cdots \partial _{\mu
_n}n^{a_n} 
\end{equation}
$$
\mu ,\mu _1,\cdots \mu _n=0,1,\cdots n;\;\;\;\;\;a_1,\cdots a_n=1,\cdots ,n 
$$
to describe the vortex state of the system and its zeroth component is
defined as the density of the total vortex charge $\rho =j^0$. In the
expression, $\partial _\mu $ stands for $\partial /\partial x^\mu $, $%
A(S^{n-1})=2\pi ^{n/2}/\Gamma (n/2)$ the area of $(n-1)$--dimensional unit
sphere $S^{n-1}$ and $n^a(x)$ is the direction field of the $n$--component
order parameter field $\vec \phi (\vec r,t)$ 
\begin{equation}
\label{3}n^a(\vec r,t)=\frac{\phi ^a(\vec r,t)}{||\phi (\vec r,t)||}%
,\;\;\;\;||\phi (x)||=\sqrt{\phi ^a(\vec r,t)\phi ^a(\vec r,t)}\ 
\end{equation}
with 
\begin{equation}
\label{2}n^a(\vec r,t)n^a(\vec r,t)=1. 
\end{equation}
It is obvious that $n^a(\vec r,t)$ is a section of the sphere bundle $S(X)$
and it can be looked upon as a map of $X$ onto $(n-1)$--dimensional unit
sphere $S^{n-1}$ in order parameter space. Clearly, the zero points of the
order parameter field $\vec \phi (\vec r,t)$ are just the singular points of 
$n^a(\vec r,t)$. From the formulas above, we conclude that in the TDGL
model, there exists a conservative equation of the topological current in (%
\ref{1})%
$$
\partial _\mu j^\mu =0,\;\;\;\;\;\mu =0,1,\cdots n 
$$
or%
$$
\frac{\partial \rho }{\partial t}+\vec \nabla \cdot \vec j=0. 
$$

In the following, we will investigate the intrinsic structure of this vortex
topological current (\ref{1}) by the use of the $\phi $--mapping method. From (\ref{2}%
) and (\ref{3}), we have 
$$
\partial _\mu n^a=\frac 1{||\phi ||}\partial _\mu \phi ^a+\phi ^a\partial
_\mu (\frac 1{||\phi ||}) 
$$
$$
\frac \partial {\partial \phi ^a}(\frac 1{||\phi ||})=-\frac{\phi ^a}{||\phi
||^3} 
$$
which should be looked upon as generalized function$^{\cite{gelfand}}$. Using
these expressions the topological current (\ref{1}) can be rewritten as%
$$
j^\mu =C_n\epsilon ^{\mu \mu _1\cdots \mu _n}\epsilon _{a_1\cdots
a_n}\partial _{\mu _1}\phi ^a\partial _{\mu _2}\phi ^{a_2}\cdots \partial
_{\mu _n}\phi ^{a_n}\frac \partial {\partial \phi ^a}\frac \partial {%
\partial \phi ^{a_1}}(G_n(||\phi ||)) 
$$
where $C_n$ is a constant%
$$
C_n=\left\{ 
\begin{array}{ccc}
-\frac 1{A(S^{n-1})(n-2)(n-1)!} & \;for & \;n>2 \\ 
\frac 1{2\pi } & \;for & \;n=2 
\end{array}
\right. 
$$
and $G_n(||\phi ||)$ is a generalized function%
$$
G_n(||\phi ||)=\left\{ 
\begin{array}{ccc}
\frac 1{||\phi ||^{n-2}} & \;for & \;n>2 \\ 
\ln ||\phi || & \;for & \;n=2. 
\end{array}
\right. 
$$
If we define $n+1$ Jacobians $J^\mu (\frac \phi x)$ as 
\begin{equation}
\label{4}\epsilon ^{a_1\cdots a_n}J^\mu (\frac \phi x)=\epsilon ^{\mu \mu
_1\cdots \mu _n}\partial _{\mu _1}\phi ^{a_1}\cdots \partial _{\mu _n}\phi
^{a_n}, 
\end{equation}
in which $J^0(\frac \phi x)$ is just the usual $n$--dimensional Jacobian
determinant%
$$
J^0(\frac \phi x)=D(\frac \phi x)=\frac{D(\phi ^1,\cdots \phi ^n)}{%
D(x^1,\cdots x^n)}, 
$$
and make use of the $n$--dimensional Laplacian Green's function relation in $%
\phi $--space$^{\cite{11}}$%
$$
\bigtriangleup _\phi (G_n(||\phi ||))=-\frac{4\pi ^{n/2}}{\Gamma (n/2-1)}%
\delta (\vec \phi ) 
$$
where $\bigtriangleup _\phi =(\frac{\partial ^2}{\partial \phi ^a\partial
\phi ^a})$ is the $n$--dimensional Laplacian operator in $\phi $--space, we
do obtain the $\delta $--function structure of the vortex topological
current rigorously 
\begin{equation}
\label{5}j^\mu =\delta (\vec \phi )J^\mu (\frac \phi x). 
\end{equation}
This expression involves the total defect information of the TDGL system and
it indicates that all of the vortices are located at the zero points of the
order parameter field $\vec \phi (\vec r,t)$. From this expression, the
density of $j^\mu $ is also changed into a compact form 
\begin{equation}
\label{6}\rho =j^0=\delta (\vec \phi )D(\frac \phi x). 
\end{equation}
We find that $j^\mu \neq 0,$ $\rho \neq 0$ only when $\vec \phi =0,$ which
is the singular point of $j^\mu $. In detail, the Kernel of $\phi $--mapping
is the singularities of the topological current $j^\mu $ in $X$, i.e. the
inner structure of topological current is labelled by the zeroes of $\phi $%
--mapping. We think that this is the core of topological current theory and $%
\phi $--mapping is the key to study it. The essential and elegance
of $\phi $--mapping theory with singularities are that the long and complex
form of the topological current $j^\mu $ in (\ref{1}) can be simply
expressed in the form of a generalized function $\delta (\vec \phi )$.

From the above discussions, we see that the Kernel of $\phi $--mapping plays
an important role in topological current theory. So we will search for the
solutions of the equations%
$$
\phi ^a(\vec r,t)=0,\;\;\;\;\;a=1,\cdots ,n 
$$
by means of the implicit function theorem, and further give the dynamic form
of the topological current $j^\mu $. Suppose the function $\phi ^a(\vec r,t)$
possesses $l$ isolated zeroes. The implicit function theorem$^{\cite{golsat}}$
says that when these zeroes are regular points of $\phi $--mapping and
require the Jacobian $D(\phi /x)\neq 0$, there is one and only one system of
continuous functions of $x^0=t$%
\begin{equation}
\label{7}\vec x=\vec z_i(t),\;\;\;\;\;i=1,\cdots ,l,
\end{equation}
which is trajectory of the $i$--th zero and is called the $i$--th
one--dimensional singular line $L_i$ in the space--time $X$. On the other
hand, putting the solutions (\ref{7}) back into $\phi ^a(x)$, we have%
$$
\phi ^a(t,\vec z_i(t))\equiv 0,\;\;\;\;\;a=1,\cdots ,n, 
$$
from which one can prove that the velocity of the $i$--th vortex is
determined by$^{\cite{golsat}}$
\begin{equation}
\label{velocity}v^\mu =\frac{dz_i^\mu }{dt}=\frac{J^\mu (\phi /x)}{D(\phi /x)%
}|_{\vec z_i(t)},
\end{equation}
taking account of (\ref{12}), the topological current (\ref{14}) can be
rewritten in a simple and compact form%
$$
j^\mu =\rho v^\mu . 
$$
It is surprised that the topological current (\ref{1}) just can take the
same form as the current density in classical electrodynamics or
hydrodynamics. The expression given by (\ref{velocity}) for the velocity is
very useful because it avoids the problem of having to specify the positions
of the vortices explicitly. The positions are implicitly determined by the
zeros of the order parameter field. The general expression with $J^\mu (%
\frac \phi x)$ should be useful in looking at the motion of vortices in the
presence of external fields beyond a growth kinetics context.

Following, we will investigate the topological charges of the vortices and
their quantization. Let $M$ be a spatial hypersurface in $X$ with variables $%
x^1,\cdots ,x^n$ for a given $t$, and $M_i$ a neighborhood of $\vec z_i(t)$
on $M$ with boundary $\partial M_i$ satisfying $\vec z_i\notin \partial M_i$%
, $M_i\cap M_j=\emptyset $. Then, the generalized Winding Number $W_i$ of $%
n^a(\vec r,t)$ at $\vec z_i(t)$ can be defined by the Gauss
map$^{\cite{Guillemin}}$ $n:$
$\partial M_i\rightarrow S^{n-1}$%
$$
W_i=\frac 1{A(S^{n-1})(n-1)!}\int_{\partial M_i}n^{*}(\epsilon _{a_1\cdots
a_n}n^{a_1}dn^{a_2}\wedge \cdots \wedge dn^{a_n}) 
$$
where $n^{*}$ is the pull back of map $n$. The generalized Winding Number is
a topological invariant and is also called the degree of Gauss map$^{\cite{19}}$.
It is well known that $W_i$ are corresponding to the first homotopy group $%
\pi [S^{n-1}]=Z$ (the set of integers)$^{\cite{gu}}$ Using the Stokes' theorem in
exterior differential form, one can deduce that%
$$
W_i=\int_{M_i}\rho d^nx. 
$$
Using the result in (\ref{6}), we get the compact form of $W_i$%
\begin{equation}
\label{8}W_i=\int_{M_i}\delta (\vec \phi )D(\frac \phi x)d^nx.
\end{equation}
Following, by analogy with the procedure of deducing $\delta (f(x))$, we can
expand the $\delta $--function $\delta (\vec \phi )$ as 
\begin{equation}
\label{9}\delta (\vec \phi )=\sum_{i=1}^lc_i\delta (\vec x-\vec z_i(t))
\end{equation}
where the coefficients $c_i$ $(i=1,\cdots ,l)$ must be positive, i.e. $%
c_i=\left| c_i\right| $. Substituting (\ref{9}) into (\ref{8}) and
calculating the integral, we get%
$$
W_i=\int_{M_i}\sum_{i=1}^lc_i\delta (\vec x-\vec z_i(t))D(\frac \phi x%
)d^nx=c_iD(\frac \phi x)|_{\vec x=\vec z_i(t)} 
$$
which gives%
$$
c_i=\frac{|W_i|}{|D(\frac \phi x)_{\vec x=\vec z_i(t)}|}. 
$$
Let $|W_i|=\beta _i$, the $\delta $--function $\delta (\vec \phi )$ can be
expressed by the zeroes of $\phi ^a(x)$ as 
\begin{equation}
\label{10}\delta (\vec \phi )=\sum_{i=1}^l\frac{\beta _i\eta _i}{D(\frac \phi
x)_{\vec x=\vec z_i(t)}}\delta (\vec x-\vec z_i(t))
\end{equation}
where the positive integer $\beta _i$ is called the Hopf index$^{\cite{20}}$ of $%
\phi $--mapping on $M_i$, the obvious meaning of $\beta _i$ is that when the
point $\vec x$ covers the neighborhood of the zero point $\vec z_i(t)$ on $M$
once, the function $\vec \phi (x)$ covers the corresponding region $\beta _i$
times. $\eta _i=signD(\phi /x)_{\vec z_i}=\pm 1$ is the Brouwer degrees$^{\cite
{20}}$ of $\phi $--mapping. The formula (\ref{10}) has the topological
information $\beta _i\eta _i$ and then is the generalization of ordinary $%
\delta $--function theory. Using this expansion of $\delta (\vec \phi )$ in (%
\ref{10}), it is evidently that the topological current $j^\mu $ in (\ref{5}%
) can be further expressed in the form 
\begin{equation}
\label{14}j^\mu =\sum_{i=1}^l\beta _i\eta _i\delta (\vec x-\vec z_i(t))\frac{%
J^\mu (\phi /x)}{D(\phi /x)},
\end{equation}
and the density of topological current $\rho $ is 
\begin{equation}
\label{12}\rho =j^0=\sum_{i=1}^l\beta _i\eta _i\delta (\vec x-\vec z_i(t)).
\end{equation}
which are exactly the current and density of a system of $l$ vortices with
topological charges $g_i=\beta _i\eta _i$ moving in the $n+1$ dimensional
space-time. The $l$ one--dimensional singular manifolds $L_i(i=1,\cdots l)$
in the space--time $X$, which are locus of the zero points $\vec z_i(t)$%
, are just the trajectory of these vortices in the space--time. The total
topological charge of this system is 
\begin{equation}
\label{13}G=\int_M\rho d^nx=\sum_{i=1}^l\beta _i\eta _i=\sum_{i=1}^lg_i.
\end{equation}

In summary, from our theory, for the first time, we obtain the topological
charges of the vortices $g_i=\beta _i\eta _i$ in the context of TDGL model
and these charges of vortices are topological quantized in terms of the Hopf
indices and the Brouwer degrees of the $\phi $--mapping. Here we see that
these vortices are located at the zeros of $\vec \phi (\vec r,t)$, i.e. the
singularities of the unit vector $\vec n(\vec r,t)$ and, the Hopf indices $%
\beta _i$ and Brouwer degree $\eta _i$ classify these vortices. In detail,
the Hopf indices $\beta _i$ characterize the absolute values of the
topological charges and the Brouwer degrees $\eta _i=+1$ correspond to
vortices while $\eta _i=-1$ to antivortices. From (\ref{6}) and (\ref{12})
the total topological charge of these vortices system in TDGL model can also
be expressed as%
$$
G=\int_M\rho d^nx=\deg \phi \int_{\phi (M)}\delta (\vec \phi )d^n\phi =\deg
\phi . 
$$
We see that the total topological charge $G$ of these system is equal to the
degree of $\phi $--mapping $\deg \phi $. And from (\ref{13}) we have the
obvious result that $\deg \phi =\sum\limits_{i=1}^l\beta _i\eta _i$, i.e.
the degree of $\phi $--mapping is equal to the sum of the indices of $n$%
--component order parameter field $\vec \phi $ at its zeros or the
topological charge of the vortices. With the discussion mentioned above, we
know that the results in this letter are obtained straightly from the
topological view point under the condition $D(\phi /x)\neq 0$. When this
condition failed, i.e. $D(\phi /x)=0$, there should exist some kinds of
branch processes in the topological current and we will discuss this problem
in other papers.

\end{document}